\newcommand{\EQ}[1] {equation~(\ref{#1})}
\newcommand{\SEC}[1] {Section~\ref{#1}}

\newcommand{\FIG}[1] {Figure~\ref{#1}}

\newcommand{\VEC}[1] {{{{\bf #1}}}}

\newcommand{\peps}{{\epsilon}}

\newcommand{\hp}{\hat{\rm \bf n}}
\newcommand{\hO}{{\hat{\rm \bf e}_{z}\,\!}}
\newcommand{\hm}{{\hat{\rm \bf e}_{x}\,\!}}
\newcommand{\hn}{{\hat{\rm \bf e}_{y}\,\!}}
\newcommand{\kg}{{k_{\rm g}}}
\newcommand{\wg}{{\omega_{\rm g}}}
\newcommand{\wc}{{\omega_{\rm cut}}}
\newcommand{\mg}{{m_{\rm g}}}

\newcommand{\debug}\bf

\documentclass[12pt]{iopart}
\usepackage{graphicx, natbib}
\begin{document}
\def\REFJNLT#1{{\rmfamily #1}}%
\newcommand\aj{\REFJNLT{AJ}}
\newcommand\araa{\REFJNLT{ARA\&A}}
\newcommand\apj{\REFJNLT{ApJ}}
\newcommand\apjl{\REFJNLT{ApJ}}
\newcommand\apjs{\REFJNLT{ApJS}}
\newcommand\ao{\REFJNLT{Appl.~Opt.}}
\newcommand\apss{\REFJNLT{Ap\&SS}}
\newcommand\aap{\REFJNLT{A\&A}}
\newcommand\aapr{\REFJNLT{A\&A~Rev.}}
\newcommand\aaps{\REFJNLT{A\&AS}}
\newcommand\azh{\REFJNLT{AZh}}
\newcommand\baas{\REFJNLT{BAAS}}
\newcommand\jrasc{\REFJNLT{JRASC}}
\newcommand\memras{\REFJNLT{MmRAS}}
\newcommand\mnras{\REFJNLT{MNRAS}}
\newcommand\pra{\REFJNLT{Phys.~Rev.~A}}
\newcommand\prb{\REFJNLT{Phys.~Rev.~B}}
\newcommand\prc{\REFJNLT{Phys.~Rev.~C}}
\newcommand\prd{\REFJNLT{Phys.~Rev.~D}}
\newcommand\pre{\REFJNLT{Phys.~Rev.~E}}
\newcommand\prl{\REFJNLT{Phys.~Rev.~Lett.}}
\newcommand\pasp{\REFJNLT{PASP}}
\newcommand\pasj{\REFJNLT{PASJ}}
\newcommand\qjras{\REFJNLT{QJRAS}}
\newcommand\skytel{\REFJNLT{S\&T}}
\newcommand\solphys{\REFJNLT{Sol.~Phys.}}
\newcommand\sovast{\REFJNLT{Soviet~Ast.}}
\newcommand\ssr{\REFJNLT{Space~Sci.~Rev.}}
\newcommand\zap{\REFJNLT{ZAp}}
\newcommand\nat{\REFJNLT{Nature}}
\newcommand\iaucirc{\REFJNLT{IAU~Circ.}}
\newcommand\aplett{\REFJNLT{Astrophys.~Lett.}}
\newcommand\apspr{\REFJNLT{Astrophys.~Space~Phys.~Res.}}
\newcommand\bain{\REFJNLT{Bull.~Astron.~Inst.~Netherlands}}
\newcommand\fcp{\REFJNLT{Fund.~Cosmic~Phys.}}
\newcommand\gca{\REFJNLT{Geochim.~Cosmochim.~Acta}}
\newcommand\grl{\REFJNLT{Geophys.~Res.~Lett.}}
\newcommand\jcp{\REFJNLT{J.~Chem.~Phys.}}
\newcommand\jgr{\REFJNLT{J.~Geophys.~Res.}}
\newcommand\jqsrt{\REFJNLT{J.~Quant.~Spec.~Radiat.~Transf.}}
\newcommand\memsai{\REFJNLT{Mem.~Soc.~Astron.~Italiana}}
\newcommand\nphysa{\REFJNLT{Nucl.~Phys.~A}}
\newcommand\physrep{\REFJNLT{Phys.~Rep.}}
\newcommand\physscr{\REFJNLT{Phys.~Scr}}
\newcommand\planss{\REFJNLT{Planet.~Space~Sci.}}
\newcommand\procspie{\REFJNLT{Proc.~SPIE}}
\let\astap=\aap \let\apjlett=\apjl \let\apjsupp=\apjs \let\applopt=\ao 

\title[Pulsar Timing Arrays and Gravity Tests in the Radiative Regime]{Pulsar 
Timing Arrays and Gravity Tests in the Radiative Regime}
\author{K J Lee$^1$}
\address{$^1$ Max-Planck-Institut f\"ur Radioastronomie, Auf dem H\"ugel 69, 
D-53121 Bonn, Germany}
\ead{kjlee007@gmail.com}
\begin{abstract}
In this paper, we focus on testing gravity theories in the radiative regime 
using pulsar timing array observations. After reviewing current techniques to 
measure the dispersion and alternative polarization of gravitational waves, we 
extend the framework to the most general situations, where the combinations of a
massive graviton and alternative polarization modes are considered.  The atlas 
of the Hellings-Downs functions is completed by the new calculations for these 
dispersive alternative polarization modes. We find that each mode and 
corresponding graviton mass introduce characteristic features in the 
Hellings-Downs function. Thus, in principal, we can not only detect each 
polarization mode, measure the corresponding graviton mass, but also 
discriminate the different scenarios. In this way, we can test gravity theories 
in the radiative regime in a generalized fashion, and such method is a direct 
experiment, where one can address the gauge symmetry of the gravity theories in 
their linearised limits.  Although current pulsar timing still lacks enough 
stable pulsars and sensitivity for such practices, we expect that future 
telescopes with larger collecting area could make such experiments be 
feasible.  \end{abstract}
\submitto{\CQG}
\maketitle

\section{Introduction}

Pulsars are compact rotating celestial objects, which radiate beamed
electromagnetic waves. Due to the rotation of pulsar, observers on
the Earth will detect pulsed emissions. Because the pulsar rotation
is highly stable \citep{VBC09}, measuring the timing residuals,
that is, the differences between observed and predicted time of
arrivals (TOAs), enables one to study the processes affecting the
timing. Gravitational waves (GWs) perturb the metric of space time,
on which the pulse propagates. In this way, one can directly detect GWs
\citep{EW75,Sazhin78,Detweiler79} using pulsar timing techniques.

In order to discriminate the GW signal from noise, one can monitor
several pulsars at the same time, compare all the TOAs, and extract
the common signal induced by the GW. Such a combination of simultaneous
pulsar timing data is called pulsar timing array (PTA), which allows
for the detection of GWs at much lower frequencies, in the nano-Hertz band.
The confidence of a positive GW detection using PTAs comes from the
fact that there will be characteristic correlations of the timing
residuals between widely-spaced pulsars, if fluctuations of TOAs are
induced by GWs. It is hard for other types of noise to mimic such spatial
correlation. The pioneering work by \cite{HD83} attempted to detect this
effect by cross-correlating the time derivative of the timing residuals
of multiple pulsars. \cite{JHLM05} improved the techniques by directly
correlating residuals, rather than their time derivative. They also
proved that detecting the stochastic GW background generated by super
massive blackhole binaries \citep{JB03, WL03,EINS04,SHMV04, SVC08, WLH09}
can be feasible with cutting-edge pulsar observing systems. Many related efforts
to detect GWs with PTA were proposed afterwards, which include various 
detection-scheduling
algorithms \citep{AB09, vHL09,BL10, YC11, CS12b, ES12, LB12}, different types of 
sources such as the single continuous sources \citep{JL04, SV08, SVV09, DF11, 
LW11, BP12, BS12, EJ12, KS12, MG12}, the burst sources \citep{FL10}, the 
stochastic background \citep{SVC08, RW12, SB12}, and the memory effects 
\citep{Seto09, HL09, CJ12}, and many other applications. Readers can consult 
the summary by \cite{Lommen12} or reviews in this volume for the related 
references.

We can study the intrinsic properties of GWs as well as the related
astrophysics via GW observations using PTAs. General relativity (GR) has
been extremely successful in describing gravitational interactions,
while there remains many feasible alternatives \citep{Will06}. Some
of these alternatives may pave the way to the final grand unification
theory of four types of fundamental interactions. In this way, testing
gravity theories is fundamentally important. In the PTA context, we
can test gravity theories in the radiative regime by investigating the
intrinsic properties of GWs \citep{LJR08, LJ10, CS12}, e.g. measuring
the polarization and dispersion of GWs. All of these efforts limit the test to 
the case, where either 1) an alternative polarization or 2) a graviton induced 
dispersion presents. 

This paper is to extended previous efforts to more general situations,
where we consider all the possible combinations of alternative GW
polarizations and gravity induced dispersion. The structure of the paper
is: In \SEC{sec:gm} and \ref{sec:pol},  we will review the current efforts
to detect the dispersion and the polarization modes of GW beyond the
prediction of GR. The extension of these work to include the more general
situation is in \SEC{sec:ext}. The conclusions and discussions are made
in \SEC{sec:con}

\section{Test gravity theories with GW dispersion relation}
\label{sec:gm}

It is well known that the corresponding particles of GWs,
as in the linearised theory of GR, have zero masses. Such zero-mass graviton
introduces an $1/r$ Newtonian potential in the static weak field limit,
and makes the phase velocity of GWs be identical to light velocity $c$
regardless of frequency. Introducing a non-zero graviton mass, in
general, leads to a theoretical issue of the so-called vDVZ discontinuity
\citep{Iwasaki70,VdV70,Zakharov70}. No matter how small the graviton
mass is, there exists a coupling of zero helicity state graviton, such
that the gravitation induced light deflection angle becomes 4/3 times that
of the GR prediction. At present, light bending measurements have
already tested GR to a very high precision.  For example, GR predicate
that the post-Newtonian parameter $\gamma=1$, and \citet{FKL09} have
confined the post-Newtonian parameter $\gamma =1 + (1 \pm 5) \times
10^{-4}$ by measuring the solar gravitational deflection of radio
waves. However, the theories involving massive gravitons cannot be
simply ruled out, because of possible mechanisms to resolve such
discontinuity \citep{Vainshtein72, Visser98, DDGV02, DKP03, FS02}. In
these mechanisms, a nonlinear contribution is usually introduced such that
the theory smoothly transits to GR at zero-mass limit \footnote{In fact,
one must include the nonlinear corrections, e.g. it is well known that
the linearised model of GR give a 4/3 times Mercury perihelion precession
rate compared to the weak field limit.}.

There are two groups of methods to probe the graviton mass. One is to
check the large scale static gravitational field, i.e. to detect the
extra exponential decay of field potential (Yukawa potential) in the form of $
e^{-r/r_{\rm
cut}}$.  Here the characteristic decay length scales $r_{\rm cut}\propto
\hbar/ m_{\rm g} c $ is inversely proportional to the graviton mass
$m_{\rm g}$, and the $\hbar$ is the reduced Planck constant. No deviation so 
far has been found in experiments performed on various scales from 
sub-millimeter
to galaxy cluster size (see \citet{GN10} for a thorough list).

Another way to probe the graviton mass is to investigate the GW
dispersion relation, i.e. the frequency dependence of the GW velocities. Such 
an approach has not been taken yet, since the direct detection of GWs has not 
yet succeeded. Because of the model dependence, one should be careful in 
interpreting the graviton mass in the two scenarios, i.e. the Newtonian and the 
linear limit. The massive GW dispersion relation can be inferred by replacing 
the corresponding terms in the particle energy-momentum relation by their
quantum versions, i.e. energy $E\rightarrow \hbar \omega_g $ and momentum 
$\VEC{p} \rightarrow \hbar \VEC{k}_{\rm g}$, where the $\omega_g$ and 
$\VEC{k}_{\rm g}$ are the angular frequency and wave vector of the GW.  The GW 
dispersion relation then becomes
\begin{equation}
 {\bf \kg}(\wg) = \frac{\left(\wg^{2}-\wc^2\right)^{\frac{1}{2}}}{c}\, 
 \hat{\VEC{n}}\,,
	\label{eq:dispf}
\end{equation}
where $\hat{\VEC{n}}$ is the unit vector in the GW propagation direction. If the
GW frequency $\wg$ is less than the cut-off frequency
$\omega_{\rm cut}\equiv\mg c^2/\hbar=3.0 \left(\frac{\mg}{10^{-23} \,
\textrm{ev}}\right) \textrm{rad}\, \textrm{yr}^{-1}$, then the wave vector 
becomes
imaginary, and the waves attenuate on the length scale of $r_{\rm cut}$. The PTA 
observations usually have cadence of more than a few days. In this way, there 
would be no GW signal in pulsar TOA, if $m_{\rm g}\ge 10^{-22}$ ev. Up to now, 
in the context of PTA, only the dispersion relations like \EQ{eq:dispf} were 
considered, although there are generalizations of this dispersion relation 
inspired by the Lorentz-violating theories or quantum theories of gravity 
\citep{MYW12}. 

The dispersion of GWs introduces a frequency dependent phase to the wave
propagation. The dispersion effect is `amplified' by the source distance,
since the phase is distance-dependent as $\VEC{k}_{\rm g}(\omega_{\rm
g})\cdot \VEC{x}$. In this way, observing the GW from a distant system and
comparing to the modeled intrinsic waveform will enable one to measure
the graviton mass, as demonstrated by \citep{Will98,AW09, SW09}. The
primary target source of PTA experiments, on the other hand, is the
stochastic GW background. In this way, the method of detecting the effect
of graviton mass using PTAs is very different from the above approach, where 
one can model the intrinsic GW waveform.

The timing residual ($R$) induced by GWs can be written as the multiplication of 
two parts. One part describes the geometrical setups of the pulsars and the 
observer on the Earth, and the other part describes the GW waveform, as shown in 
the following equation, one can check \cite{LJ10} for the detailed calculations.
\begin{equation}
	R=-\frac{1}{\cal S} A^{ij} H_{ij},\label{eq:res}
\end{equation}
The part $H_{ij}$ describes the GW amplitude and propagation such that
\begin{equation}
	H_{ij}=  \int_{0}^{\tau}h_{ij}(\tau,0)-h_{ij}(\tau-
 |{\rm \bf D}|/c,{\rm \bf D} )\,d\tau,
 \label{eq:H}
\end{equation}
and $\cal S$ involves the dispersion relation of the GW, \begin{equation}
	{\cal S}=2\left (1+({c}/{\wg}){\rm \bf \kg} \cdot \hp\right )
\end{equation}
The part that describes the geometrical setups is \begin{equation}
	A^{ij} \equiv \hp^{i}\hp^{j}. \label{eq:geoten}
\end{equation}
Here the observer is at the coordinate origin, $\rm \bf D$ is the displacement 
vector from the observer to the pulsar, $\hp^{i}$ is the direction from the 
observer (Earth) to the photon's source (pulsar). One can see that the spatial 
correlation of pulsar timing residuals, i.e. the cross-power $C$ between two 
different pulsars, is $C_{1,2}(\theta)=\langle R_{1} R_{2}\rangle ={A}_1{A}_2 
\langle {\cal S}_1 {\cal S}_2 {H}_1 {H}_2\rangle$, where the sub-scripts are 
indexes for the pulsars, the $\theta$ is the angle between the two pulsars.  
Clearly, the graviton mass introduces its signature in the spatial correlation 
of pulsar timing residuals via $\cal S$ and $H$, since both depend on the 
dispersion relation of the GW. 

We now try to understand the effects of the graviton mass in terms of such
cross-power. From \EQ{eq:H}, one can see that \begin{equation}
	\lim_{\omega_{\rm g}\rightarrow \omega_{\rm cut}} R= 
	A^{ij}\frac{1}{2}\int_{0}^{\tau} h_{ij}(\tau,0)-h_{ij}(\tau-|\VEC{D}|/c, 
	0)d\tau.
\end{equation}
At such a limit, the cross power become $C\propto {A}_1 {A}_2 {h}
{h}$, i.e. it is the power of the projected GW strain tensor $h_{ij}$ onto the 
basis spanned by the pulsar directions. In this situation, one can expected 
that $C(\theta)$ will be of $90^{\circ}$ symmetry for the two 
transverse-traceless modes predicted by GR, i.e. the quadruple nature of the 
$h_{ij}$ gives $C(90^\circ-\theta)=C(90^\circ+\theta)$.  Apart from this limit, 
the $90^{\circ}$ symmetry $h_{ij}$ will not manifest in the final correlation 
function. This is mainly because the term $\cal S$, describing the wave 
propagation, breaks the symmetry around $90^\circ$.

\section{Test gravity theories with GW polarimetry}
\label{sec:pol}

Besides the dispersive properties, the other type of intrinsic properties
helping to test gravity theories is the allowed polarization modes of
the GW.  It is well known that GR predicts two allowable modes, the
transverse-traceless modes. One would expected that, since gravity is
mediated via the spin-2 particles, the maximal number of polarization
modes of GWs would be five, i.e.  $2s+1=5$, and $s=2$. However,
one can prove \citep{ELL73, LJR08} that six polarization modes are
required to fully describe the most general polarization states of
GW. The possible states are: 1) two helicity-0 modes, a transverse
breathing and a longitudinal modes, 2) two helicity-1 shear modes, 3)
two transverse-traceless modes. Such apparent incompatibility between
the five spin states and six polarization modes lies in the assumption
of the two analysis frameworks. The spin classification is based on the
unitary presentation of the wave-function transforms, while the analysis
for GWs was not limited to the case \footnote{The presentation of the little 
group for the GW transformation here, i.e. the subgroup of the Lorentz group 
keeping GW wave 4-vector invariant, is indecomposable \citep{Dirac84} in the 
vector space spanned by the six polarization modes \citep{ELL73}. Note that 
there is a typo in the definition of ``indecomposable'' in that paper.}. 

The cross power between the timing residual of pulsar pairs depends on
the angular separation of the pulsar. The so-called Hellings-Downs (HD)
function is defined as the cross-correlation coefficient of the pulsar
timing residuals, i.e. the cross power normalized by the signal power.
The shape of HD function, taking pulsar angular separation as the
argument, is sensitive to the polarization state of GWs. In this way, one can 
detect and differentiate the GW polarization modes by measuring the HD 
function.  The way that
GW polarization enters the HD function can be seen from \EQ{eq:res},  where the 
$H_{ij}$ is polarization dependent. 

The power of pulsar timing residuals also critically depends on the GW 
polarization. Many authors \citep{LJR08, AT11, CS12} noticed that the timing 
residuals induced by longitudinal GWs will be amplified by the distance from 
observer to the pulsars. \citet{CS12} further found that certain pulsar pairs 
with very small angular distances can also be used to increase the sensitive of 
detecting longitudinal modes.  The two shear modes share similar features of 
the longitudinal mode, although the amplifying factor is rather limited due to 
its logarithmic dependence on the pulsar-Earth distances.  For the rest of the 
modes, i.e. all transverse modes, the pulsar timing responses are similar and 
nearly independent of the pulsar-Earth distance. One can understand such 
distance dependence by considering a limiting case, where the GW propagates 
along the line of sight of the pulsar. For the longitudinal modes, the GW and 
photon travel in phase, and there is a non-oscillatory contribution to the 
photon's red shift. Such a non-oscillatory contribution accumulates and gives 
an amplifying factor proportional to the pulsar distance. 

The above analysis is all based on the assumption that the GW travels at
the same speed as the photons. However, one would naturally expect that
the alternative polarization modes could also be dispersive, especially
because introducing a graviton mass would break the gauge symmetries and
thus allow for those alternative polarizations to be excited. Previously,
there were investigation on the pulsar timing response and cross-power for
either 1) massive GR modes, or 2) massless alternative modes. To
fully understand the influence of GWs on pulsar timing residuals,
it is desirable to extend current works to the more general case. In
next section, we will try to complete this mission by studying the pulsar
timing response to these massive alternative modes.

\section{Pulsar timing response to massive GW stochastic background with 
alternative polarization modes}
\label{sec:ext}

The HD functions for a stochastic GW background with massive gravitons are 
affected by the total length of the data and sampling duration \citep{LJ10}. It 
seems to be impossible to analytically calculate the HD function for the 
stochastic background with a power law spectrum \footnote{One can analytically 
calculate the HD function for the monochromatic component of the stochastic 
background.  However the results are too lengthy to be really practically 
useful.}.  Thus, we will calculate the Hellings-Downs function via numerical 
simulations, while the induced signal power, i.e.  $C(\theta=0)$, is calculated 
analytically to guide the physical intuition.

For a stochastic GW background, these metric perturbations
are a superposition of monochromatic GWs with random phase and amplitude,
and can be written as \begin{equation}
	h_{ij}(t,r^i)=\sum_{P}\int_{-\infty}^{\infty}df_{\rm g}\int
	d\Omega\, h^{P}(f_{\rm g},\hO )\, \peps^{P}_{ij}(\hO) e^{ i[\wg
	t-{\bf k}_g(\wg)\cdot{\bf r }} ], \label{eq:grbk}
\end{equation} where $f_{\rm g}={\wg}/{2\pi}$ is the GW frequency,
$\Omega$ is solid angle, spatial indices $i,j$ run from 1 to 3,
and $h^{P}$ is the amplitude of the GW propagating
in the direction of $\hO$ per unit solid angle, per unit frequency
interval, in polarization state $P$, i.e. the six states $+, \times,
b, l,sn, se$ defined in \citet{LJR08}.  The basis tensors for
these polarization are $\epsilon^{P}_{ij}$ are \begin{eqnarray}
\peps^{ +}_{ab}&=&\hm_{a}\hm_{b}-\hn_{a}\hn_{b}, \nonumber\\
\peps^{ \times}_{ab}&=&\hm_{a}\hn_{b}+\hn_{a}\hm_{b}, \nonumber\\
\peps^{b}_{ab}&=& \hm_{a}\hm_{b}+\hn_{a}\hn_{b},\nonumber\\ \peps^{
sn}_{ab}&=& \hm_{a}\hO_{b}+\hO_{a}\hm_{b},\nonumber\\ \peps^{
se}_{ab}&=& \hn_{a}\hO_{b}+\hO_{a}\hn_{b},\nonumber\\ \peps^{ l}_{ab}&=&
\hO_{a}\hO_{b}\,.  \label{eqpoltensor}\end{eqnarray} If the GW background is 
isotropic, stationary and independently polarized,
we can define the characteristic strain $h_{c}^{P}$ according to
\begin{equation} \langle h^{P}(f,\hO)
h^{P'\star }(f',\hO') \rangle=\frac{1}{4\pi}\delta(f-f')\delta(\hO-\hO
')\delta_{PP'}\frac{|h_{\rm c}^{P}|^2}{\eta(P)f}\,.  \label{eq:hcdef}
\end{equation} where $\star$ stands for the complex conjugate
and $\langle \rangle$ is the statistical ensemble average. The
symbol $\delta_{PP'}$ is the Kronecker delta for polarization states;
$\delta_{PP'}=0$ when $P$ and $P'$ are different, and $\delta_{PP'}=1$,
when $P$ and $P'$ are the same. If we define $\eta(P)\equiv4$,
for $P=+,\times,b,\rm sn,se$, and $\eta(P)\equiv2$, for $P=l$, one can show that 
\begin{equation}
	\langle
	h_{ab}(t)h_{ab}(t)\rangle=\sum_{P=+,\times}\int_{0}^{\infty}\frac{|h_{c}^{P}|^{2}}{f_{\rm
	g}} df_{\rm g}\,.
\end{equation}

\begin{figure}[ht]
	 \includegraphics[totalheight=5.5in]{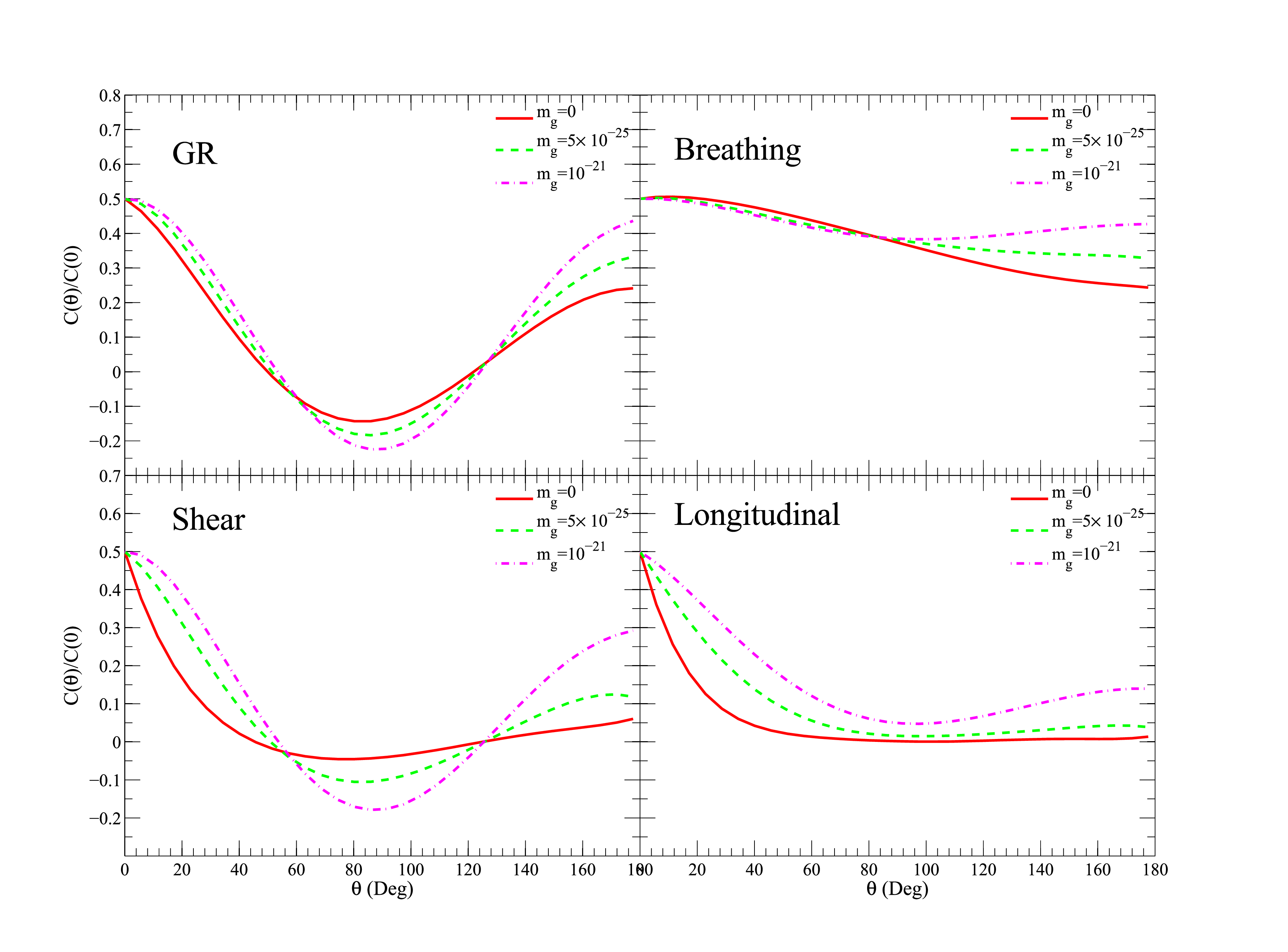}
	\caption{The HD correlation functions of all six polarization modes. The 
	solid, dashed, and dot-dashed curves are for graviton mass of $0$, $5\times 
	10^{-25}$, and $10^{-22}$ ev respectively. In the calculation, we assume that 
	the total	observing length is 10 years, while the cadence is two weeks. The 
	distance of the pulsars are all assumed to be 1 kpc. One can clearly see the 
	graviton mass dependence of the HD function of each polarization modes.}
		\label{fig:hel}
\end{figure}

We use Monte-Carlo simulations to determine the shape of the
HD function for GW backgrounds with a power-law spectra. In the simulations for
$C(\theta)$, we randomly choose pulsars from an isotropic distribution
over sky positions. We then hold constant these pulsar positions
and calculate the angular separation $\theta$ between every pair of
pulsars. Next, to simulate the power-law GW background, we generate
$10^4$ monochromatic waves, choosing random phase, and choosing the
amplitude according to the power-law such that the characteristic strain
$h_{\rm c}=A_{\rm c} (f/f)^{\alpha}$ , where we take $\alpha= −2/3$ for
practical purposes \citep{Phinney01}. The timing residuals are calculated using
\EQ{eq:res}. Then, the cross power, $C(\theta)$, between
pulsar pairs is calculated.  We repeat such processes and average over
the angular dependent cross-power $C$ until its change is less than 0.1\%. From 
cross power $C(\theta)$, we calculate the Hellings-Downs curve according to 
$H(\theta)=C(\theta)/C(0)$, where $C(0)$ is just the power of the GW induced 
signal of single pulsar.  The Hellings-Downs curves for all six polarization 
with different graviton masses are plotted in \FIG{fig:hel}.

We now analytically calculate the power spectrum of pulsar timing residuals 
induced by the stochastic GW background. One can show that power spectrum of 
the induced pulsar timing residuals is
\begin{equation}
	S_{\rm R}(f_{\rm g})=\int d\Omega \frac{ |h_{\rm c}^{P}|^2}{16 \pi f_{\rm g}} 
	\left[\frac{\epsilon_{ij}^{P}A^{ij}}{\omega_{\rm g}+c{\rm \bf \kg}\cdot \hp} 
	\right]^2 \left[1-\cos \left(\frac{D}{c} (\omega_{\rm g}+ c{\rm \bf \kg}\cdot 
	\hp)\right)\right]\,.
	\label{eq:spec}
\end{equation}
After integrating out the solid angle $d\Omega$ in \EQ{eq:spec}, one can show 
that \footnote{These results are partly due to the independent polarization 
assumption, which assumes no correlation between the polarizations via 
$\delta_{PP'}$ as in \EQ{eq:hcdef}. However only an extra term involving the 
longitudinal and breathing modes, the $\langle h^{b} h^{l}\rangle$, will 
appear, if we further allow for the cross-correlation of the modes, i.e. the 
`circular' polarization. Interestingly, the correlation function and power of 
such term are identical to that of shear modes.}
\begin{equation}
	S_{\rm R}(f)=\sum_{P} \frac{|h_{\rm c}^{P}|^2}{24\pi^2 f^3} F^{P}.
\end{equation}
Here the $F^{P}$ is the amplifying factor of the pulsar timing response to the 
polarization mode $P$ normalized to the massless GR case. The amplifying 
factors depend on the mass of graviton and the pulsar distances as
\begin{eqnarray}
	F^{+,\times}=\int_{-1}^{1} \frac{3 \left(1-\mu ^2\right)^2 \sin 
	^2\left[\frac{1}{2} (\zeta  \mu  \Phi +\Phi )\right]}{4 (\zeta \mu +1)^2}\, 
	d\mu ,\label{eq:fpt}\\
	F^{b}=2F^{+,\times}, \label{eq:fb}\\
	F^{sn,se}=\int_{-1}^{1} \frac{3 \mu ^2 \left(1-\mu ^2\right) \sin 
	^2\left[\frac{1}{2} (\zeta  \mu  \Phi +\Phi
	)\right]}{(\zeta  \mu +1)^2}\,d\mu, \label{eq:fen}\\
	 F^{l}=\int_{-1}^{1} \frac{3 \mu ^4 \sin ^2\left[\frac{1}{2} (\zeta  \mu  \Phi 
	 +\Phi )\right]}{(\zeta  \mu +1)^2}\,d\mu,\label{eq:fl}
\end{eqnarray}
where $\zeta\equiv\sqrt{1-(\omega_{\rm cut}/\omega_{\rm g})^2}$, and $\Phi=D 
\omega_{\rm g} /c$ is the pulsar distance in unit of GW wave
length\footnote{$\Phi$ is just the pulsar distance. It is not the phase 
difference between the pulsar and earth term, which is $\Delta \Phi= D/c 
(\omega_{\rm g} +c \VEC{k}_{\rm g} \cdot \hp)$. }. Although the amplifying 
factor can be integrated analytically (given in the \ref{sec:appf}), the exact 
form is complex and awkward to use. Here we check the behavior of $F^{P}$ in 
two limiting case $\zeta=0$ and $\zeta=1$. The Cases $\zeta=1$ corresponds to 
the massless graviton, while $\zeta=0$ is the situation where the GW wave 
vector vanishes, i.e. the corresponding graviton momentum is zero. One can show 
that \begin{eqnarray}
	F^{+,\times}=\left\{\begin{array}{c c}
 1+{\cal O}\left(\frac{1}{\Phi^2}\right)&
 \textrm{when } \zeta=1+{\cal O}\left(\frac{1}{\Phi^2}\right), \\
 \frac{2}{5}\left(1-\cos\Phi\right), & \textrm{when } \zeta=0,
	\end{array}\right. \label{eq:fptl}\\
	F^{b}=2F^{+,\times},\\
	F^{sn/se}=\left\{\begin{array}{c c}
 3 \log \Phi +3 \gamma -7+\log 8+{\cal O}\left(\frac{1}{\Phi}\right)&
\textrm{when } \zeta=1, \\
\frac{2}{5}\left(1-\cos(\Phi)\right), & \textrm{when } \zeta=0,
	\end{array}\right. \\
	F^{l}=\left\{\begin{array}{c c}
\frac{3 \pi  \Phi }{4}-6 \log \Phi -6 \gamma +\frac{37}{4}-3 \log 4 +{\cal 
O}\left(\frac{1}{\Phi}\right), &
\textrm{when } \zeta=1, \\
\frac{3}{5} \left(1-\cos\Phi\right), & \textrm{when } \zeta=0,
	\end{array}\right. \label{eq:fll}
\end{eqnarray}
where the symbol $\cal O$ means ``the order of''. Here we assumes the short-wave 
approximation that $\Phi\gg1$, i.e. the pulsar distance is much larger than the 
zero mass wavelength of GW. Such an assumption is valid for most PTA 
applications, where $\Phi\simeq 10^{3}$.

One interesting effect is that the cross power of pulsar pairs separated by 
$180^\circ$, the $C(180^\circ)$, also gain an exta distance dependent 
amplification
for the longitudinal mode. One can show that
\begin{equation}
	C^{l}(180^\circ)=\frac{|h_{\rm c}^{P}|^2}{192 \pi^2 f^3}\left[ 3 \log \Phi +3 
	\gamma -8+\log 8\right].
\end{equation}
However, due to the logarithmic dependence on $\Phi$, the effect of boosting the 
PTA sensitivity is rather limit compared to the case of $\theta\simeq 0$.

Another thing worth mentioning is that the Hellings and Downs curves have very 
similar shape at the cut-off frequency limit, i.e. $\zeta=0$. In this case, the 
integrations in \EQ{eq:fpt}$-$\EQ{eq:fl} are very simple, and we can derive the 
analytic form that
\begin{eqnarray}
	C^{+,\times}(\theta)/C^{+,\times}(0)=\frac{1}{4}\left(1+3\cos2\theta\right), 
	\\
	C^{b}(\theta)/C^{b}(0)=\frac{1}{8}\left(7+\cos2\theta\right), \\
	C^{sn,se}(\theta)/C^{sn,se}(0)= C^{+,\times}(\theta)/C^{+,\times}(0), \\
	C^{+,\times}(\theta)/C^{+,\times}(0)=\frac{1}{3}\left(2+\cos2\theta\right).
\end{eqnarray}
All the HD function are practically identical to each other, since they only 
differ by constant offsets, to which PTAs are not sensitive, because of the 
clock noise. Thus, discrimination between the polarization modes could be 
difficult, if the graviton mass is not negligibly small.

\section{Discussion and Conclusion}
\label{sec:con}

In this paper, we reviewed the current efforts to test alternative gravity
theories by measuring the mass of the graviton and detecting alternative
polarization modes of GWs, in the context of PTA experiments. We
explained the method to measure the graviton mass or detect alternative
GW polarization by measuring the cross-correlation function of pulsar
timing residuals, i.e. HD functions. This method is different from the
techniques for single GW sources, where one can model the waveform and
compare it to the observations. We do not need
to model the waveform for these PTA experiments. Thus the gravity test will
be independent of the models of  GW sources or generation mechanisms.

We have also extended previous calculations for the Hellings-Downs function to 
the most general case, where one allows for the alternative polarization modes 
as well as dispersion due to a massive graviton. Such the extension is driven 
by the consideration that alternative polarization modes usually originate
from gauge symmetry breaking, which could be induced by a graviton mass.  
Because of such connections between the massive graviton and polarization 
modes, it is necessary to consider both effects at the same time, when trying 
to test gravity theories using PTA
data. 

One particular interesting case is the longitudinal modes. It is known that an 
large amplifying factor, e.g. the $3\pi\Phi/4$ in \EQ{eq:fll}.  Most of the 
pulsars in PTA has distance of a few thousand light years. This makes the 
amplifying factors be about $10^4$. Several authors \citep{LJR08, AT11, CS12} 
have realized this could be a very good opportunity to put tight bounds on the 
amplitude of longitudinal modes. However, these modes are likely to be a massive 
modes and, according to \EQ{eq:fll}, the amplifying factor drops to order of 
unity, when approaching the cut-off frequency.  Thus, a null detection of 
longitudinal modes should not be just interpreted as the upper bounds for the GW 
amplitude.  The null detection can be also be explained by a small graviton 
mass.  A similar effect is seen for the shear modes.  However the effect is not 
that dramatic, beucase the amplifying factor is proportional to the logarithmic 
of the pulsar distance, which only increases the pulsar timing response by a 
order of magnitude. The drop-off of the amplifying factor by the graviton mass 
is natural. When the GW frequency approaches the cut-off frequency, the wave 
vector vanishes and the wavelength becomes arbitrarily large. This effectively 
stops the spatial oscillation of the GW, reduces the difference between pulsar 
and Earth term, and brings down the amplifying factors.

We have shown that the Hellings-Downs function is not a single function. They 
are, in fact, six sets of functions depending on the polarization and graviton 
mass. In this paper, we did not go into the details of the related signal 
detection and discrimination problems, as they beyond the scope of this paper.  
On the other hand, from simulations, forty to a few hundred pulsars are 
required to differentiate the difference in Hellings-Downs functions via a 
correlation analysis \citep{LJR08, LJ10}. In this way, testing gravity theories 
using PTAs would be, unfortunately, beyond current ability of PTAs. However, 
the number of pulsars ($N$) one can time to a given accuracy is proportional to 
the fourth power of the telescope effective diameter according to the 
radiometer equation.  In this way, we would expect that future telescopes with 
larger effective diameters, such as the Large European
Array for Pulsars (\citealt{BVK09}, Kramer \& Champion,2013, in this volume), 
the Five-hundred-meter Aperture
Spherical Radio Telescope \citep{NWZZJG04, SLKMSJN09} and the Square
Kilometer Array (SKA, Lazio, 2013, in this volume) will offer unique 
opportunities to find and time these pulsars; and to detect the GW background 
and test gravity theories in the radiative regime. 

\ack
KJL gratefully acknowledges support from the ERC Advanced Grant
`LEAP', Grant Agreement Number 227947 (PI Michael Kramer).  The author thank 
Lijing Shao, C.~M.~F. Mingaralli, and members in the EPTA data analysis group 
for useful discussions. The author also thank D.~Champion, M.~Kramer for 
reading the manuscript and for their valuable comments.  \label{lastpage}
\clearpage
\appendix
\section{The full form of amplifying factors}
\label{sec:appf}
Integration of \EQ{eq:fpt}$-$(\ref{eq:fl}) can be perform analytically using the 
Sine-Cosine integrals, where one has
\begin{eqnarray}
F^{+,\times}&=&\frac{W}{4 \zeta ^5}+\frac{3 \cos \Phi  \sin (\zeta  \Phi )}{2 
\zeta ^5 \Phi ^3}-\frac{3
	 (\sin \Phi \sin (\zeta  \Phi )+\zeta  \cos \Phi \cos (\zeta  \Phi ))}{2 \zeta 
	 ^5 \Phi
	 ^2}\nonumber \\
	 &+&\frac{3 \left(\zeta ^2-3\right) \cos \Phi \sin (\zeta  \Phi )+6 \zeta  
	 \sin \Phi  \cos
	 (\zeta  \Phi )}{4 \zeta ^5 \Phi }\nonumber \\
	 &+&\frac{3 \left(\zeta ^2-1\right)^2 \Phi  [\textrm{Si}(\zeta  \Phi
	 +\Phi )-\textrm{Si}(\Phi -\zeta  \Phi )]}{8 \zeta ^5},\\
F^{b}&=&2F^{+,\times}, \\
	F^{sn, se}&=&	\frac{V}{\zeta ^5}-\frac{6 \cos \Phi  \sin (\zeta  \Phi
	 )}{\zeta ^5 \Phi ^3}+\frac{6 (\sin \Phi  \sin (\zeta  \Phi )+\zeta  \cos \Phi 
	 \cos (\zeta \Phi ))}{\zeta ^5 \Phi ^2}\nonumber \\
	 &+&\frac{9 \cos \Phi  \sin (\zeta  \Phi )-6 \zeta  \sin \Phi \cos
	 (\zeta  \Phi )}{\zeta ^5 \Phi }\nonumber \\
	 &+&\frac{3 \left(\zeta ^2-1\right) \Phi  [\textrm{Si}(\zeta  \Phi +\Phi
	 )-\textrm{Si}(\Phi -\zeta  \Phi )]}{2 \zeta ^5}, \\
	F^{l}&=&\frac{U}{2 \zeta ^5 \left(\zeta ^2-1\right)}+\frac{3
	 \cos \Phi  \sin (\zeta  \Phi )}{\zeta ^5 \Phi ^3}-\frac{3 (\sin \Phi  \sin 
	 (\zeta  \Phi
	 )+\zeta  \cos \Phi  \cos (\zeta  \Phi ))}{\zeta ^5 \Phi ^2}\nonumber \\
	 &+&\frac{6 \zeta  \sin \Phi  \cos
	 (\zeta  \Phi )-3 \left(\zeta ^2+3\right) \cos \Phi  \sin (\zeta  \Phi )}{2 
	 \zeta ^5 \Phi
	 }\nonumber \\
	 &+&\frac{3 \Phi  [\textrm{Si}(\zeta  \Phi +\Phi )-\textrm{Si}(\Phi -\zeta  
	 \Phi )]}{4 \zeta ^5},
\end{eqnarray}
Here the $U,V$, and $W$ are defined as \begin{eqnarray}
	U&=&6 \left(\zeta ^2-1\right) [\textrm{Ci}(\zeta  \Phi +\Phi 
	)-\textrm{Ci}(\Phi -\zeta  \Phi )]\nonumber\\
&+&\zeta^5+8 \zeta ^3-12 \left(\zeta ^2-1\right) \tanh ^{-1}\zeta+3 \sin \Phi  
\sin (\zeta  \Phi) \nonumber\\
&+&3\zeta  \cos \Phi  \cos (\zeta  \Phi )-12 \zeta,\\
V&=&3 \left(\zeta ^2-2\right) [\textrm{Ci}(\zeta  \Phi +\Phi )-\textrm{Ci}(\Phi 
-\zeta  \Phi )]+2 \zeta
		\left(\zeta ^2-6\right) \nonumber\\
		&-&6 \left(\zeta ^2-2\right) \tanh ^{-1}\zeta +3 \sin (\Phi ) \sin (\zeta 
		\Phi )+3 \zeta  \cos (\Phi ) \cos (\zeta  \Phi ), \\
		W&=&6 \left(\zeta ^2-1\right) [\textrm{Ci}(\Phi -\zeta  \Phi 
		)-\textrm{Ci}(\zeta  \Phi +\Phi )]+3
	 \left(\zeta ^2-1\right) \sin \Phi  \sin (\zeta  \Phi )\nonumber \\
	 &+&4 \left(-2 \zeta ^3+3 \left(\zeta ^2-1\right) \tanh ^{-1}\zeta+3
	 \zeta \right)\nonumber \\
	 &+&3 \zeta  \left(\zeta ^2-1\right) \cos
	 \Phi  \cos (\zeta  \Phi ).
\end{eqnarray}

These equations are the exact calculation for the amplification factors, i.e. no 
assumption such as $\Phi\gg1$ are made. In this way, these equation also valid 
to calculate the timing residual with the much lower frequencies compared to the 
case of \EQ{eq:fptl}$-$(\ref{eq:fll}).

\end{document}